## MOND reveals the thermodynamics of gravity

Peter V. Pikhitsa

Seoul National University, 151-742, Korea

E-mail: peter@snu.ac.kr

**Abstract**

We show that treating gravitation as a thermodynamical theory leads to the modified Newton dynamics (MOND) equations if one takes into account the Hubble's expansion. Then the universal MOND acceleration $a_0$ is exactly twice the product of the light velocity $c$ and the Hubble constant $H$. No dark matter is needed for the description of the galaxy rotational curves as well as for the accounting for the additional gravitational lensing at large distances.

### 1 Introduction

Long ago Milgrom introduced in astrophysics an *ad hoc* theory that could explain the rotation of galaxies without the hypothetical dark matter on cost of modification of Newton dynamics (MOND) at accelerations less than some universal value $a_0 \approx 10^{-8}$ cm/s$^2$ [1]. Later on Milgrom guessed that the origin of MOND could lie in the thermodynamics of the vacuum as seen by accelerating body [2] due to Davies-Unruh effect.

We show that this guess is fruitful but the reasoning of [2] should be applied to the gravitational field. In that way we can remove the current ambiguity between two possible MONDs – kinetics modified [1] and the potential modified theories [3, 4] at least for the circular motion. We also demonstrate that the additional gravitational lensing naturally takes place in MOND without complicated field theories like TeVeS (**Te**nsor-**Ve**ctor-**S**calar) [4] which actually (by introducing hypothetical fields) smuggle the dark matter through the back door. In light of recent formulation of gravity as a thermo-dynamical effect [5, 6] we confirm that there is no dark matter needed to explain the properties of galaxies and that the origin of MOND lies in the thermodynamics of the gravity. In what follows we will use the units $\hbar = c = k_B = 1$.

## 2 Gravitation in the expanding Universe

We take into account the presence of the Hubble constant $H$ which can be interpreted as a curved space-time with the homogenous curvature $K$ [7]:

$$K = H^2 . \tag{1}$$

Let us show now that the gravity field is modified when vacuum effects are taken into account to obtain the constant velocity of the galaxy rotation curves with the correct estimation for $a_0 \approx H \approx 10^{-8}$ cm/sec$^2$ and the correct gravitation lensing effect. Indeed, the surface gravity is [5, 6]

$$g^{\nu} = -e^{\phi} \nabla^{\nu} \phi \tag{2}$$

and therefore the motion in the gravity field should be described by its potential exactly like in Newton dynamics only with the redshift factor for the stationary case [8]. We will take the limit of low accelerations and low potentials where the exponential factor can be neglected. In case of energy equipartition over the number of states $dn$ the thermo-dynamical equation for the gravity produced by a mass $M$ reads [5]:

$$E = M = 2 \int_{\partial V} T dS = \frac{1}{2} \int_{\partial V} T dn \quad . \tag{3}$$

Here the integration is over the surface $\partial V$ and it is always possible for a stationary case to choose the surface located at equipotential surfaces or “holographic screens” [6], the local temperature is $T = g/2\pi$ and $g = n_\nu \nabla^\nu \phi = \sqrt{\left(\vec{\nabla}\phi\right)^2} = \left|\vec{\nabla}\phi\right|$, where $n_\nu$ is the unit vector normal to the surface element $dA$ so that $dA^\nu = n^\nu dA$; the entropy of the surface element is $dS = dA/4L_P^2 = dA/4G$, where $L_P$ is the Planck length, $G$ is the gravity constant. The Newton law of gravitation follows directly from (3) [5, 6].

In the curved space-time under consideration there is the Davies-Unruh temperature [9] $T_0 = K^{1/2}/2\pi = H/2\pi$ in the absence of the mass *M*. This temperature should be taken into account in (3) by noticing that actually introducing the mass *M* should not change the entropy of the space-time thus the mass can be introduced adiabatically. Then the local temperature alternates [10]:

$$T = \sqrt{T_0^2 + \frac{g^2}{4\pi^2}} = \frac{1}{2\pi}\sqrt{g^2 + H^2} \tag{4}$$

and according to the first law of thermodynamics instead of (3) we obtain

$$2\int_{\partial V} T_0 dS + M = 2\int_{\partial V} T dS \tag{5}$$

Finally, equation (5) can be rewritten as

$$\int_{\partial V}\left(\sqrt{g^2 + H^2} - H\right) dA \frac{1}{4\pi G} = M \tag{6}$$

This equation can also be written as

$$\int_{\partial V} \mu(g/H) \nabla^\nu \phi \, dA_\nu = 4\pi G M \tag{7}$$

with

$$\mu(x) = \sqrt{1+\frac{1}{x^2}} - \frac{1}{x} \tag{8}$$

Therefore we obtained nearly the general MOND version suggested in [3] (see Eq. (27) therein) but without any TeVeS-like constructions [4] although the formula (8) for $\mu(x)$ is very close to the one predicted in [4]; in fact, (8) coincides with the one suggested in [2] heuristically if one introduces Milgrom's acceleration constant $a_0 = 2H$. Indeed, at the large distance $R$ when $x << 1$ and the spherical symmetry one obtains from (7) for the surface gravity the MOND expression

$$\frac{g^2}{a_0} = \frac{GM}{R^2} \tag{9}$$

instead of the Newtonian one for $x >> 1$

$$g = \frac{GM}{R^2} \; . \tag{10}$$

As far as for such a faraway circular orbit one has to equate the circular acceleration $a = \frac{V^2}{R}$ to $g$ from (9) one obtains the final result for the rotation velocity $V = (GMa_0)^{1/4}$ [11]. In a general case of arbitrary motion one may use the equation

$$a^\nu = g^\nu = -\nabla^\nu \phi \tag{11}$$

as suggested in [3] and define $\phi$ from (6) or (7). These equations can be rewritten in a local form as a modified Poisson equation [3] with the mass density $\rho_M$ :

$$\vec{\nabla}\left[\mu\left(\frac{|\vec{\nabla}\phi|}{H}\right)\vec{\nabla}\phi\right] = 4\pi\rho_M \; . \tag{12}$$

Gravitational lensing is easily predicted as far as instead of the Newtonian potential one gets the logarithmic potential from (9) in the spherically symmetric case which has been successfully imitated by the linear distribution of the dark matter mass [12]. The deflection angle of the light between two light asymptotes should reproduce the one calculated for the dark matter case in [12]:

$$\Delta\alpha = 2\pi V^2. \tag{13}$$

Indeed, by slightly modifying expressions for the light deflection angle from [13] one may write down a general expression in the spherically symmetric case:

$$\Delta\alpha = 4\rho\int_{\rho}^{\infty}\frac{g\,dR}{\left(R^2-\rho^2\right)^{1/2}}\ , \tag{14}$$

where $\rho$ is the distance of the closest approach of the light to the center. In case of Newton gravity (10) this leads to a well known result $\Delta\alpha = 4GM/\rho$ and for the modified gravity (9) the calculation gives exactly (13). Again no dark mass is needed.

## 3 Conclusions

We demonstrated that MOND is a rigorous physical theory which turned out to be a revelation of thermodynamics responsible for gravity. As far as the Hubble's constant is involved in the MOND structure making the dark matter not needed in Physics, one may expect an interesting turn connected the problem of the $\Lambda$ - constant and the dark energy as well.

It would be interesting to follow the similar arguments as given in [5, 6] for possible inertia modification. One would expect a deviation from the Newtonian inertia similar to originally introduced by Milgrom [1] and it would involve only the acceleration along the trajectory. For example, one can notice that for the circular motion when the acceleration is orthogonal to the velocity there is no work produced therefore the thermodynamics of space-time may predict no deviation from the Newtonian inertia.

After the first version of the present paper was submitted to the ArXiv I was shown an article that reported similar thermodynamic arguments supporting the MOND to eliminate the dark matter from the galaxies [14]. This paper also delves into Cosmology (see [15] as well). Another interesting paper [16] gives an insight for a possible fine structure of the holographic screen and interprets the MOND as a quantum statistics phenomenon away from the equipartition law. One can notice that the main difference between our still equipartition equations (5) , (6) and the corresponding low temperature Debye-function equation in [16] lies in the difference in the definition of the temperature at the holographic screen: in our interpretation the temperature cannot be lower than $H$ so the equipartition seems to be a reasonable approximation.